\documentclass[10pt,aps,prd,preprintnumbers,superscriptaddress,showpacs,
twocolumn,
floatfix,nofootinbib]{revtex4-2}
\usepackage{graphicx}
\usepackage{amsmath}
\usepackage{slashed}
\usepackage{xcolor}

\newcommand{\be}{\begin{equation}}
\newcommand{\ee}{\end{equation}}

\allowdisplaybreaks

\newcommand{\ep}{\epsilon}
\newcommand{\ci}{\mathrm{i}}

\newcommand{\PROD}[2]{\prod\limits_{#1}^{#2}}
\newcommand{\ORD}[1]{\mathop{}\!\mathcal{O}\!\left(#1\right)}

\newcommand{\INT}[2]{\int_{#1}^{#2}}

\newcommand{\DIFFL}{\mathop{}\!\mathrm{d}}

\newcommand{\scr}[2]{(#1\!\cdot\! #2)}

\newcommand{\DD}[1]{\mathop{}\!\delta\!\left(#1\right)}

\newcommand{\HT}[1]{\mathop{}\!\theta\!\left(#1\right)}

\newcommand{\GF}[1]{\mathop{}\!\Gamma\!\left(#1\right)}
\newcommand{\GFP}[2]{\mathop{}\!\Gamma^#1\!\left(#2\right)}
\newcommand{\GENHYPGF}[5]{\mathop{}\!_{#1}F_{#2}\!\left[\{#3\},\{#4\};#5\right]}
\newcommand{\HYPGF}[4]{\GENHYPGF{2}{1}{#1,#2}{#3}{#4}}

\newcommand{\xLi}[2]{\mathop{}\!{\mathrm{Li}}_{#1}\!\left(#2\right)}

\newcommand{\Ln}[1]{\mathop{}\!{\mathrm{ln}}(#1)}

\newcommand{\Lnp}[2]{\mathop{}\!{\mathrm{ln}}^{#1}(#2)}

\newcommand{\MZV}[1]{\mathop{}\zeta_{#1}}

\newcommand{\CFcl}[2]{\mathop{}\!{\mathrm{Cl}}_{#1}\!\left(#2\right)}
\newcommand{\CFclp}[3]{\mathop{}\!{\mathrm{Cl}}_{#1}^{#2}\!\left(#3\right)}

\begin{document}


\title{\boldmath
  Same-hemisphere three-gluon-emission contribution to the zero-jettiness soft function at N3LO QCD
  \unboldmath}

\author{Daniel Baranowski}
\email[]{daniel.baranowski@kit.edu}
\affiliation{Institut f\"ur Theoretische Teilchenphysik,
 Karlsruher Institut f\"ur Technologie (KIT), 76128 Karlsruhe, Germany}

\author{Maximilian Delto}
\email[]{maximilian.delto@tum.edu}
\affiliation{Physik Department, Technische Universit\"at M\"unchen, James-Franck-Strasse 1, 85748, Germany}

\author{Kirill Melnikov}
\email[]{kirill.melnikov@kit.edu}
\affiliation{Institut f\"ur Theoretische Teilchenphysik,
  Karlsruher Institut f\"ur Technologie (KIT), 76128 Karlsruhe, Germany}

\author{Chen-Yu Wang}
  \email[]{chen-yu.wang@kit.edu}
  \affiliation{Institut f\"ur Theoretische Teilchenphysik,
  Karlsruher Institut f\"ur Technologie (KIT), 76128 Karlsruhe, Germany}

  \begin{abstract}
    We complete the calculation of the three-gluon-emission contribution to the same-hemisphere  part 
  of  the zero-jettiness soft function at next-to-next-to-next-to-leading order
  in perturbative QCD. 
 \end{abstract}

\preprint{P3H-22-039,TTP22-025,TUM-HEP-1396/22}

\maketitle

\section{Introduction}
\label{sect:intro}
The goal of this paper is to present the result for the same-hemisphere three-gluon-emission contribution to
the zero-jettiness soft function at next-to-next-to-next-to-leading order in perturbative QCD.
Computation of such corrections needs to overcome  several technical challenges that were discussed in 
  Ref.~\cite{Baranowski:2021gxe}.  These challenges stem  from the fact 
 that an  observable that defines the soft function --  the
 so-called jettiness --  involves  Heaviside functions. These Heaviside functions are needed to distinguish
 between emissions of
 soft gluons into two hemispheres,  defined relative  to directions of incoming hard radiators. 

 Presence of  Heaviside functions complicates  application of generalized unitarity~\cite{Anastasiou:2002yz}
 and  integration-by-parts identities~\cite{Chetyrkin:1981qh}
 to phase-space integrals.
 We have  discussed in Ref.~\cite{Baranowski:2021gxe} how to overcome this problem and  explained  
 how to derive useful integration-by-parts relations
for integrals with Heaviside functions.  To show the efficacy of this method, 
we   employed   an eikonal function derived in Ref.~\cite{Catani:2019nqv}, which  describes emissions of
three soft gluons,  and
integrated it over soft-gluon phase space subject to
zero-jettiness constraints. We restricted ourselves to  contributions
where  all gluons are emitted into the same hemisphere. 

Unfortunately, in Ref.~\cite{Baranowski:2021gxe},
we have not  completed the  computation of this ``same-hemisphere'' contribution. 
Indeed,
the representation  of the eikonal function derived in Ref.~\cite{Catani:2019nqv}
involves four terms  and in Ref.~\cite{Baranowski:2021gxe} we have fully  integrated three of them.
The fourth contribution  can be written in the following way 
\begin{align}
S_d = \int {\rm d} \Phi^{nnn}_{\theta \theta \theta} \, \omega^{(3),d}_{n\bar{n}}(k_1,k_2,k_3) \,,
\label{eqn:s_d_def}
\end{align}
where $k_{1,2,3}$ are four-momenta of final-state gluons,  $\Phi^{nnn}_{\theta \theta \theta}$ is the phase space subject
to zero-jettiness conditions, cf. Eq.~(\ref{eq2a}), 
and $\omega^{(3),d}_{n\bar{n}}(k_1,k_2,k_3) $ is the eikonal function defined in
Eq.~(5.49) in Ref.~\cite{Baranowski:2021gxe}.  We note that the integral in Eq.~(\ref{eqn:s_d_def}) is the 
the most complicated  one,  as it contains a propagator that
depends on the  relative orientation of four-momenta of all three soft gluons.

Although we described  a possible
way to calculate this contribution in  Ref.~\cite{Baranowski:2021gxe}, we did not complete its  computation there.
The goal of this paper  is to compute the missing piece
and to present the result for the same-hemisphere three-gluon-emission contribution to the zero-jettiness soft function.

The rest of the paper is organized as follows. In  Section~\ref{sect:intwd}  we explain the general
strategy that we used to integrate the
function $\omega_{n\bar{n}}^{3,(d)}$. In Section~\ref{sect:intJbar2nu}
we describe the  computation of the boundary condition for one of the master integrals
that provides a contribution that has unusual sensitivity to an analytic regulator.
In Section~\ref{sect:4} we discuss  checks that we performed to ensure the correctness of the
computation.  In Section~\ref{sect:5} we present the result for the same-hemisphere three-gluon-emission contribution to the
zero-jettiness soft function  and  conclude in Section~\ref{sect:6}. 

\section{Integrating  $\omega_{n\bar{n}}^{3,(d)}$}
\label{sect:intwd}

The missing part of the  same-hemisphere contribution to the zero-jettiness soft function, displayed in Eq.~(\ref{eqn:s_d_def}),
requires integration of the function $\omega_{n\bar{n}}^{3,(d)}$,
which contains terms with a  propagator $1/k_{123}^2$, where
$k_{123}=k_1+k_2+k_3$.

To integrate   the function $\omega_{n\bar{n}}^{3,(d)}$, we consider a class of integrals
\begin{align}
    I_{\theta \theta \theta}
    & =
    \int \; {\rm d} \Phi^{n n n}_{\theta \theta \theta} \;
    \frac{(k_1 n)^\nu (k_2 n)^\nu (k_3 n)^\nu}{k_{123}^2 \; (k_{1} k_{2}) (k_{1} k_{3}) \cdots },
    \label{eq1}
\end{align}
where $n$ and $\bar n$ are two light-like four-vectors pointing in the direction of incoming partons, ellipses stand for eikonal propagators,\footnote{These are all possible scalar products $q\!\cdot\!v$, where $q$ is a linear combination of four-vectors of soft gluons and $v$ is one of the two light-cone vectors $n$ or $\bar n$.} and ${\rm d} \Phi^{n n n}_{\theta \theta \theta}$ is the normalized phase-space measure. It is defined as follows~\cite{Baranowski:2021gxe}
\be
    {\rm d} \Phi^{n n n}_{f_{1} f_{2} f_{3}}
    =
    N_{\ep}^{-3}
    \left( \prod_{i = 1}^{3} [{\rm d} k_{i}] f_{i}(k_{i}) \right)
    \delta\left( 1 - \sum_{i = 1}^{3} k_i n \right)
    ,
\label{eq2a}
    \ee
where the normalization factor is
\be
    N_{\ep}
    =
    \frac{\Omega^{(d-2)}}{4 (2\pi)^{d-1}}
    =
    \frac{(4 \pi)^{\ep}}{16 \pi^{2} \Gamma(1 - \ep)}
    ,
\ee
the function $f_i$ is either $\theta(k_i \bar n - k_i n)$ or $\delta(k_i \bar n - k_i n) $,
and
\be
    [{\rm d} k_i]
    =
    \frac{{\rm d}^d k}{(2\pi)^d} 2 \pi \delta(k_i^2) \theta(k_{i}^{0}).
    \ee
We note that we set the jettiness variable $\tau$ to one since the final result is a uniform function of $\tau$. 
We also note that in the spirit of integration-by-parts, all propagators that appear
in the integrand in Eq.~(\ref{eq1})  can be raised to arbitrary integer powers.

Finally, as can be seen from  Eq.~(\ref{eq1}), we introduced scalar products of
the gluon four-momenta $k_{1,2,3}$ with the light-cone vector $n$ raised to power $\nu$
into the integrand of $I_{\theta \theta \theta}$.   As we explained in Ref.~\cite{Baranowski:2021gxe} they are required
because some integrals, that appear in the course of the IBP reduction,  contain divergences that are not regulated
dimensionally; the  analytic regulator $\nu$ is introduced to regulate them.
To simplify the notation, we define the regulated phase-space measure to be
\be
    {\rm d} \Phi^{\nu}_{f_{1} f_{2} f_{3}}
    =
    {\rm d} \Phi^{n n n}_{f_{1} f_{2} f_{3}} (k_1 n)^\nu (k_2 n)^\nu (k_3 n)^\nu
    .
\ee

Unfortunately, even after integrals defined in
Eq.~(\ref{eq1}) are reduced to master integrals, the master
integrals with $1/k_{123}^2$ propagators appear to be too complicated
for direct analytic integration.  For this reason, as explained in Ref.~\cite{Baranowski:2021gxe}, we  derive
differential equations satisfied by these integrals,  and  solve them numerically.
To  do that, we introduce a mass parameter into the propagator that contains the momenta of all three gluons\footnote{Similar ideas that use auxiliary mass-scales have been presented, for example, in Refs.~\cite{Henn:2013nsa,Papadopoulos:2014lla,Liu:2017jxz,Lee:2022nhh}.}
\be
\frac{1}{k_{123}^2} \to \frac{1}{k_{123}^2 + m^{2}}.
\label{eq2}
\ee
The appearance  of the mass parameter $m$ allows us to
differentiate with respect to it and use integration-by-parts to derive differential equations for relevant integrals.
We then  fix  boundary conditions at $m \to \infty$, solve differential equations {\it numerically}
and determine relevant integrals at $m=0$.
This can be done by matching the numerical solution with the formal solution at $m = 0$ and then taking the $m \to 0$ limit
in an appropriate manner.
Since, as we explained in Ref.~\cite{Baranowski:2021gxe},
the differential equations can  be solved to an arbitrary precision as a matter of principle, 
and to more than 2000 digits in practice,  we  have used the
high-precision numerical results for master integrals
to find   the analytic form of the solution by  fitting 
them to  a basis of transcendental constants and rational numbers. 

To reiterate,  once it is understood how to use
generalized unitarity to write down integration-by-parts identities for integrals with $\theta$-functions,
it becomes  a fairly standard problem  to derive differential equations for relevant integrals.
However, great care is needed when 
choosing the basis of master integrals because of the analytic regulator $\nu$; in essence,
we need to find a basis that admits a simple  $\nu \to 0$ limit.
Ideally, this  should happen   for master integrals that appear in the soft function 
$S_d$, as well as  in $m$-dependent differential equations that we need to solve. 

To find a suitable basis, we use the following consideration.
In spite of the fact that the reduction to master integrals
performed {\it after} setting $\nu=0$ in Eq.~(\ref{eq1}) is incorrect, it  gives us a good idea about integrals that are independent 
of each other  in the $\nu \to 0$ limit.
Hence, to find a suitable basis for master integrals, we start by performing  a  reduction
of integrals shown in Eq.~(\ref{eq1}) at $\nu = 0$.  We then insist
that the master integrals found in the course of such a reduction should be also
chosen as master integrals for the reduction
of integrals with  $\nu \ne 0$, to an extent possible.
We do this  for integrals with and without auxiliary  mass parameter $m$. With this choice of master integrals  we find that the integral of the soft function
$\omega_{n\bar{n}}^{3,(d)}$ written in
terms of master integrals,
as well as the differential equations that these integrals obey,
admit a simple $\nu \to 0$ limit.

For example the result of the reduction of $S_d$ to master integrals can be written
in the  following way
\be
S_d
= \sum \limits_{\alpha}^{} c_\alpha(\nu) I_\alpha^\nu + \nu \sum \limits_{\alpha} \tilde c_\alpha(\nu) {\bar I}_\alpha^\nu \,,
\label{eq2.5}
\ee
where the coefficients $c_\alpha$ and $\tilde c_\alpha$ are  regular in the $\nu \to 0$ limit. The list of integrals $\{I_\alpha^\nu\}$ coincides with the
list of master integrals for $S_d$ that one obtains performing a reduction at $\nu=0$.  New integrals that
appear in the reduction, which we denote as ${\bar I}_\alpha^\nu$ in the above equation, 
are multiplied with the parameter $\nu$ and, therefore, 
disappear if the naive $\nu \to 0$ limit is taken.

However,  the naive $\nu \to 0$ limit in Eq.~(\ref{eq2.5}) cannot be taken because some of ${\bar I}_\alpha^\nu$
integrals are $1/\nu$ divergent and, therefore, need to be retained.
Examples of such integrals without the $1/k_{123}^2$ propagator can be found in Ref.~\cite{Baranowski:2021gxe}.

Unfortunately, many integrals among  $I_\alpha^\nu$ and $ {\bar I}_\alpha^\nu$ contain the $1/k_{123}^2$ propagator; to study these integrals,
we use    Eq.~(\ref{eq2}) and turn them  into  integrals with the mass parameter $m$.
We construct a system of differential equations w.r.t.\ the mass parameter $m$ by including  every integral with
the $1/k^2_{123}$ propagator  that appears in Eq.~(\ref{eq2.5}) and other integrals that
are needed to close it.
Such a system  takes the following form
\be
\begin{split} 
  & \frac{\partial}{\partial m^{2}} \boldsymbol{J}^\nu = \boldsymbol{M}_{1}(\nu) \boldsymbol{J}^\nu + \nu \boldsymbol{N}_{1}(\nu) \boldsymbol{\bar J}^\nu,\\
  & \frac{\partial}{\partial m^{2}} \boldsymbol{\bar J}^\nu = \boldsymbol{M}_{2}(\nu) \boldsymbol{\bar J}^\nu + \boldsymbol{N}_{2}(\nu) \boldsymbol{J}^\nu.
\end{split} 
\label{eq2.6}
\ee
In Eq.~(\ref{eq2.6}) matrices $\boldsymbol{M}_{1, 2}$ and $\boldsymbol{N}_{1, 2}$
are regular in the $\nu \to 0$ limit, $\boldsymbol{J}^\nu$ integrals are the
master integrals that need to be considered for computing $\{I_{\alpha}^{\nu}\}$ integrals
and $\boldsymbol{\bar J}^\nu$ are  integrals that are needed
for computing $\{{\bar I_{\alpha}}^{\nu}\}$ integrals.

From the structure of differential equations in Eq.~(\ref{eq2.6}) it is clear that taking the limit $\nu \to 0$
at the early stages of the computation is beneficial since this makes the system of differential equations significantly
simpler.  However, this is only possible if we know which master integrals are singular in the $\nu \to 0$ limit and which
master integrals are not.
Unfortunately, it is not trivial to answer this question and  we use several approaches to clarify it. 

First, from the computation of integrals of the eikonal functions
$\omega^{a,b,c}$ reported in Ref.~\cite{Baranowski:2021gxe},
we know $1/\nu$-divergent integrals that appear in cases when the propagator $1/k_{123}^2$ is absent.
Upon inspection, we find that the only $m$-independent
$1/\nu$-divergent integral that appears in the amplitude $S_d$
and in the differential equations reads
\begin{equation}
    {\bar I}_{1}^{\nu} = {\bar J}_1^{\nu}
     =
    \int \frac{\mathrm{d} \Phi_{\theta \delta \theta}^{\nu}}{(k_{1} \cdot k_{3}) (k_{1} \cdot n) (k_{1 2} \cdot \overline{n}) (k_{3} \cdot \overline{n})} \,.
\end{equation}
This integral was computed  in Ref.~\cite{Baranowski:2021gxe} and  for this reason  we do not discuss  it here.

Second, to determine which of the integrals  with  the propagator $1/k_{1 2 3}^{2}$ are singular in the $\nu \to 0$ limit,
we can study these integrals at {\it finite} values of $m$ and then employ  differential equations to determine
the $1 / \nu$ behavior of the corresponding ${\bar I}_{\alpha}^{\nu}$ integrals.
To this end,  we employed Mellin-Barnes representation of the relevant
integrals and used public programs \texttt{MB}~\cite{Czakon:2005rk} and \texttt{MBresolve}~\cite{Smirnov:2009up}
to numerically compute all $m$-dependent integrals that appear in the differential equations at finite
values of $m$.  We also used the program \texttt{pySecDec}~\cite{Borowka:2017idc,Borowka:2018goh} for cross-checks of the numerical computation. 
Upon doing that, we discovered yet another integral that  is singular in the $\nu \to 0 $ limit.  It reads 
\begin{align}
    {\bar J}_{2}^{\nu}
    =
    \int
    \frac{\mathrm{d} \Phi_{\theta \delta \theta}^{\nu}}{(k_{1 2 3}^{2} + m^{2}) (k_{1} \cdot k_{3}) (k_{1} \cdot n) (k_{1 2} \cdot \overline{n})} \,.
\end{align}
This integral is quite  peculiar and we explain in the next section
why this is the case and how to compute it. 

To proceed further,  we re-scale $\boldsymbol{\bar J}^\nu$ integrals that appear in Eq.~(\ref{eq2.6}) 
by a factor $\nu$.  Since, as we just explained,  only two integrals  
${\bar J}_{1}^{\nu}$ and ${\bar J}_{2}^{\nu}$ diverge in the $\nu \to 0$ limit, we need to keep them
in $\ORD{\nu^0}$ part of the differential equations.  Therefore, we combine
$\boldsymbol{J}^\nu$ integrals together with $\nu {\bar J}_{1}^{\nu}$ and $\nu {\bar J}_{2}^{\nu}$
integrals into a vector  $\boldsymbol{\mathcal{J}}^{\nu}$ and use remaining ${\bar J}^\nu$ integrals, rescaled
by a parameter $\nu$, to define a new vector $\boldsymbol{\mathcal{ \bar J}}^{\nu}$
\begin{align}
\begin{split}
    \boldsymbol{\mathcal{J}}^{\nu}
    & =
    \left( \boldsymbol{J}^{\nu}, \nu \bar{J}_{1}^{\nu}, \nu \bar{J}_{2}^{\nu} \right)
    ,
    \\
    \boldsymbol{\mathcal{\bar{J}}}^{\nu}
    & =
    \left( \nu \bar{J}_{3}^{\nu}, \cdots \right)
    .
\end{split}
\end{align}
We then obtain a new system of differential equations 
\begin{align}
\begin{split} 
  & \frac{\partial}{\partial m^{2}} \boldsymbol{\mathcal{J}}^\nu
  = \boldsymbol{\cal M}_{1}(\nu) \boldsymbol{\cal J}^\nu +  \boldsymbol{{\cal N}}_{1}(\nu) \boldsymbol{\bar {\cal J}}^\nu,\\
  & \frac{\partial}{\partial m^{2}} \boldsymbol{\bar {\cal J}}^\nu = \boldsymbol{{\cal M}}_{2}(\nu) \boldsymbol{\bar {\cal J}}^\nu
  + \nu \boldsymbol{\cal N}_{2}(\nu) \boldsymbol{{\cal J}}^\nu.
\end{split} 
\label{eq2.9}
\end{align}
It is straightforward to solve the above equation expanding in $\nu\to0$ because all the matrices that
appear there have smooth $\nu \to 0$ limits and because
the integral vectors
satisfy $\boldsymbol{\mathcal{J}}^\nu \sim \ORD{\nu^0}$,
$\boldsymbol{\mathcal{\bar J}}^\nu \sim \ORD{\nu}$.

Working to order $\ORD{\nu^0}$, we can drop $\boldsymbol{\mathcal{\bar J}}^\nu$ integrals and set $\nu$ to zero
in Eq.~(\ref{eq2.9}). This leads to a significant reduction
in the number of integrals that appear in Eq.~(\ref{eq2.9}) 
and allows for solving the system of differential equations in a more efficient way.


%

As  follows from the differential equation satisfied by $\nu {\bar J}_{1}^{\nu}$ and $\nu {\bar J}_{2}^{\nu}$,
these integrals are independent of $m$ through $\ORD{\nu^0}$. In addition, they
also have to obey the following relation 
\be
    \lim_{\nu \to 0} \nu {\bar J}_{1}^{\nu}
    =
    \frac{1 + 6 \ep}{1 + 4 \ep} \;
    \lim_{\nu \to 0} \nu {\bar J}_{2}^{\nu}
    ,
    \label{eq2.10}
    \ee
to cancel the $1/\nu$ singularity which  is naively  present in the  rescaled differential equation. 
We will show in the following section  that  this  condition is  indeed satisfied.

We note that the above discussion applies to differential equations at \emph{finite} values of $m$ whereas, eventually, we are
interested in the solutions at $m=0$. This limit is, potentially, non-trivial. Indeed, to find required values
of integrals at $m=0$,  we need to compute master integrals using the following sequence of limits:
$m\to 0,\nu\to0,\ep\to0$.

However, as explained above,  we would like
to simplify differential equations by taking the $\nu\to0$ limit \textit{first} and
there are two problems that may arise if the order  of limits is changed. 
First,  since the mass parameter $m$ can \emph{also} serve as a regulator of
collinear and soft singularities, $m \to 0$ and $\nu \to 0$ limits should not necessarily commute.
Second, additional  contributions  can mix into the Taylor $m^0$-branch of the solution that we require,
if the $\nu \to 0$ limit is taken before the $m \to 0$ limit. 
We will discuss  these two problems now.

Suppose we take the $\nu \to 0$ limit for finite $m$-integrals but the resulting integrals are still not regulated
dimensionally at $m=0$. This feature can  be detected in the following way. The dependence of the integral
on $m$ and $\ep $ at small $m$ has the following form 
\be
\mathcal{J} \sim \sum \limits_{n_1,n_2,n_3}^{}  c_{n_1 n_2 n_3 } m^{n_1 + n_2 \ep} \ln^{n_3}(m).
\label{eq2.12}
\ee
We are interested in taking the $m \to 0$ limit of this solution at fixed $\ep$. However, this is only possible
if the coefficients $ c_{n_10n_3} $ with $n_1 < 0$ and $n_1 = 0, n_3 > 0$
vanish so that there are no  $1/m$ and $\log m$ terms that are not multiplied by
additional powers of $m^\ep$ or sufficiently high  powers of $m$.    We have checked that this condition is satisfied for all
$\nu = 0$ integrals that we considered; this implies that the $m \to 0$ limit does not lead to divergencies in integrals 
that are not regulated dimensionally. It follows that indeed ${\bar I}_{1, 2}^{\nu}$ are the only integrals in ${\bar I}_{\alpha}^{\nu}$ that contribute to the amplitude, all other integrals can be safely discarded. 

The second problem concerns possible mixing between different branches of  integrals if the $\nu \to 0$ limit
is taken too early. To see how this comes about, consider 
a  general solution in the limit $m\to0,\nu\to0,\ep\to0$
\be
\mathcal{J} \sim \sum \limits_{n_1,n_2,n_3,n_4}^{}  c_{n_1 n_2 n_3 n_4 } m^{n_1 + n_2 \ep + n_4 \nu } \ln^{n_3}(m).
\label{eqn:ansatzm0_full}
\ee
 If there are
terms that correspond to $n_1 =0, n_2 =0, n_3 = 0$, $n_4 \neq 0$, they will mix with
the contribution $n_1 = 0, n_2 = 0, n_3 = 0, n_4 = 0$, i.e.~the $m^0$-branch that we are interested in.

Hence, we need to understand
if such contributions exist and, if they are there, how to isolate and remove them. This can be done by studying exact differential equation Eq.~(\ref{eq2.9}) at small values of $m$ but with full
$\nu$- and $\epsilon$-dependence, and checking
if $m^{n_{4} \nu}$ solutions without additional dependencies of exponents on $\epsilon$ and
additional powers of $m$ are possible.
We find that this does not happen at $m = 0$ and, therefore, the Taylor branch does not receive
any unwanted contributions.

In summary, we can solve the differential equations Eq.~(\ref{eq2.9}) as an expansion in $\nu$. 
As discussed in Ref.~\cite{Baranowski:2021gxe}, we can compute the  boundary conditions at $m = \infty$
and then find values of integrals $\boldsymbol{\mathcal{J}}$ at $m=0$ by discarding all  terms that
have non-analytic dependencies on $m$ at $m=0$.  We present the results of such a computation in
Section~\ref{sect:5}; in the next section we describe computation of a peculiar boundary condition
which illustrates that our worry about potential mixing of a Taylor- and $m^{n_4\nu}$-branches  is not unfounded.

\section{Integral ${\bar J}_{2}^{\nu}$ with $1/\nu$ divergence}
\label{sect:intJbar2nu}

We can illustrate some points discussed in the previous section
by considering the integral ${\bar J}_{2}^{\nu}$ and its contribution
to differential equations. As mentioned  there, this integral is  
singular in the limit $\nu \to 0$.   Multiplied by  a factor $\nu$, it appears 
in the differential equations for two sets  of $\nu$-regular integrals. We will consider one
of them for the sake of example
\begin{align}
J_{a_{1}}^{\nu}
& =
\int
\frac{\mathrm{d} \Phi_{\theta \delta \theta}^{\nu}}{(k_{1 2 3}^{2} + m^{2}) (k_{1} \cdot k_{3}) (k_{2 3} \cdot \bar{n})} \,,
\nonumber \\
J_{a_{2}}^{\nu}
& =
\int
\frac{\mathrm{d} \Phi_{\theta \delta \theta}^{\nu}}{(k_{1 2 3}^{2} + m^{2}) (k_{1} \cdot k_{3}) (k_{2 3} \cdot \bar{n})^{2}} \,.
\label{eq14a}
\end{align}
To compute these integrals
from differential equations, we require
a boundary condition for $\nu {\bar J}_2^\nu$, which we compute at  $m = \infty$.
It follows from the analysis of the differential equation for $\nu {\bar J}_2^\nu$ at $m = \infty$
that  the required boundary condition can be determined from the leading, mass-independent term in the  $1/m$-expanison.
Upon inspecting the various contributions to the asymptotic behaviour
of the integral ${\bar J}_2^{\nu}$ discussed in Ref.~\cite{Baranowski:2021gxe}, we find that they do not
produce 
mass-independent terms that are $1/\nu$-divergent.

It turns out that  the integral ${\bar J}_2^\nu$ provides an example of a situation
where the analysis of different regions that contribute to $m \to \infty$
asymptotic behaviour of integrals, performed in Ref.~\cite{Baranowski:2021gxe},  is \textit{incomplete} 
and that there is another  region  that needs to be considered. 
In fact we have found that the   following scaling of integration variables
\begin{align}
    & k_{3} \cdot \bar{n} = \alpha_3
    \sim
    m^{2}
    \gg
    1,
    \nonumber
    \\
    \label{eqn:new_region}
    & k_1  \cdot \bar n = \alpha_1 \sim 1,
    \\
&  
    k_{1} \cdot n = \beta_1 
    \sim
    m^{-2}
    \ll
    1,
    \nonumber
\end{align}
leads to a $1/\nu$-divergent $\ORD{1/m^0}$ contribution to ${\bar J}_2^\nu$.

To show this, we write an approximation to the integrand
of   ${\bar J}_2^{\nu}$-integral in the region defined by Eq.~(\ref{eqn:new_region})
\begin{widetext}
\begin{align}
     & {\bar J}_{2}^{\nu}
    \sim  \widetilde{J_{2}^{\nu}}
     = \int \! \frac{\mathrm{d} \Phi_{\theta \delta \theta}^{\nu}}{( \scr{k_3}{\bar{n}} \, \scr{k_2}{n} + m^{2})  \scr{k_1}{k_3}  \scr{k_1}{n}  \scr{k_{12}}{\bar{n}} } \,.
\end{align}
\end{widetext}

We use the Sudakov variables 
$\alpha_{i}$ and $\beta_{i}$ (see Ref.~\cite{Baranowski:2021gxe} for details)
and  take into account the condition  $\beta_1 \ll 1 $  to remove $\beta_1$ from the
``jettiness'' delta-function $\delta(1- \beta_{1} - \beta_{2} - \beta_{3}  ) \to \delta(1-\beta_{2} - \beta_{3}) $.
We then extend the integration over $\beta_1$ to infinity. We find
\begin{widetext}
\begin{align}
        \widetilde{J_{2}^{\nu}} & = 
   2 \int \limits_{0}^{\infty} {\rm d} \beta_1 \int \mathrm{d}  \beta_{2} \mathrm{d}  \beta_{3} \;
    \mathrm{d} \alpha_{1} \mathrm{d} \alpha_{3} \;
    \beta_{1}^{- \ep + \nu}
    \beta_{2}^{- 2 \ep + \nu}
    \beta_{3}^{- \ep + \nu}
    \alpha_{1}^{- \ep}
    \alpha_{3}^{- \ep}
    \;
    \frac{\delta(1 - \beta_{2 3}) \theta(\alpha_{1} - \beta_{1}) \theta(\alpha_{3} - \beta_{3})}{(\alpha_{3} \beta_{2} + m^{2}) \beta_{1} (\alpha_{1} + \beta_{2})}
     \nonumber \\
     \label{eqn:j2nutilde_1}
    & \phantom{= {}}
    \times
    \Bigg [
        \frac{\theta(\beta_{1} / \alpha_{1} - \beta_{3} / \alpha_{3})}{\beta_{1} \alpha_{3}}
             {}_{2}F_{1} \left( 1, 1 + \ep; 1 - \ep; \frac{\alpha_{1} \beta_{3}}{\alpha_{3} \beta_{1}} \right)
            +
        \frac{\theta(\beta_{3} / \alpha_{3} - \beta_{1} / \alpha_{1})}{\beta_{3} \alpha_{1}}
        {}_{2}F_{1} \left( 1, 1 + \ep; 1 - \ep; \frac{\alpha_{3} \beta_{1}}{\alpha_{1} \beta_{3}} \right)
    \Bigg ] \,.
\end{align}
We change integration variables $\alpha_1 = \beta_1/\xi_1$ and ${\alpha_3 = \beta_3/\xi_3}$ and obtain
\begin{align}
        \widetilde{J_{2}^{\nu}}
        & =
        2
        \int \mathrm{d} \beta_{2} \mathrm{d} \beta_{3} \mathrm{d} \xi_{1} \mathrm{d} \xi_{3}
        \beta_{2}^{- 2 \ep + \nu}
        \beta_{3}^{- 2 \ep + \nu}
        \xi_{1}^{\ep - 1}
        \xi_{3}^{\ep - 1}
        \,
        \frac{\delta(1 - \beta_{2 3}) \theta(1 - \xi_{1}) \theta(1 - \xi_{3})}{\beta_{3} \beta_{2} + m^{2} \xi_{3}}
        \nonumber \\
        \label{eqn:j2nutilde_2}
        & \phantom{= {}}
        \times
        \left(
            \xi_{3} \theta(\xi_{1} - \xi_{3})
            \HYPGF{1}{1+\ep}{1-\ep}{\frac{\xi_{3}}{\xi_{1}}}
            +
            \xi_{1} \theta(\xi_{3} - \xi_{1})
            \HYPGF{1}{1+\ep}{1-\ep}{\frac{\xi_{1}}{\xi_{3}}}
        \right) \;
        \\
        & \phantom{= {}}
        \times
        \int \limits_{0}^{\infty} \mathrm{d} \beta_{1}
        \frac{ \beta_{1}^{- 2 \ep + \nu - 1} }{\beta_{1} + \beta_{2} \xi_{1}} \,. \nonumber
\end{align}
\end{widetext}

Integrating over $\beta_1$, we find
\begin{align}
        \int \limits_{0}^{\infty} \mathrm{d} \beta_{1}
        \frac{\beta_{1}^{- 2 \ep + \nu - 1} }{\beta_{1} + \beta_{2} \xi_{1}}
        & =
        (\beta_{2} \xi_{1})^{- 2 \ep + \nu - 1}
        \Gamma(- 2 \ep + \nu)
        \nonumber \\
        & \phantom{= {}}
        \times
        \Gamma(2 \ep - \nu + 1).
\end{align}

We use this result in Eq.~(\ref{eqn:j2nutilde_2}), change variables
$\xi_{1} = r \xi_{3}$, and arrive at
\begin{align}
    \widetilde{J_{2}^{\nu}}
    & =
    2
    \Gamma(- 2 \ep + \nu)
    \Gamma(2 \ep - \nu + 1)
    \nonumber \\
    & \phantom{= {}}
    \times
    \int \mathrm{d} \beta_{2} \mathrm{d} \beta_{3}
    \beta_{2}^{- 4 \ep + 2 \nu - 1}
    \beta_{3}^{- 2 \ep + \nu}
    \delta(1 - \beta_{2 3})
    \\
    & \phantom{= {}}
    \times
    \int_{0}^{1} 
    \frac{\mathrm{d} \xi_{3} \xi_{3}^{\nu - 1} J(\nu, \xi_{3})}{\beta_{3} \beta_{2} + m^{2} \xi_{3}} \nonumber \,,
\end{align}
where
\begin{align}
    J(\nu, \xi_{3})
    & =
    \int_{0}^{1} \mathrm{d} r
    \left[
        \theta(1 - \xi_{3} / r)
        r^{\ep - \nu}
        +
        r^{- \ep + \nu - 1}
    \right]
    \nonumber \\
    & \phantom{= {}}
    \times
    \HYPGF{1}{1+\ep}{1-\ep}{r} \,.
\end{align}
It follows from the above expression that the  $\nu$-pole originates from a singularity at  $\xi_3 = 0$.
It is straightforward to compute it
since $J(\nu,\xi_3)$ is  regular at $\xi_3 = 0$.\footnote{As with any analytic regulator, the $\nu \to 0$ limit should be computed keeping $\ep$ fixed.} We find
\begin{align}
    \int_{0}^{1} \mathrm{d} \xi_{3}
    \frac{\xi_{3}^{\nu - 1} J(\nu, \xi_{3})}{\beta_{3} \beta_{2} + m^{2} \xi_{3}}
    =
    \frac{J(0, 0)}{\nu}
    (\beta_{3} \beta_{2})^{-1}
    +
    \mathcal{O}(\nu^{0}) \,.
\end{align}

Upon further integration, we obtain  a $1/\nu$-divergent contribution to $\tilde J_2^\nu$. It reads
\begin{align}
    \widetilde{J_{2}^{\nu}}
    =
    \frac{C_{2}}{\nu}
    ,
\end{align}
where
\begin{widetext}
\begin{align}
    C_{2}
    ={} &
    \frac{2 \Gamma^{2}(-2 \ep) \Gamma(-4 \ep - 1) \Gamma(2 \ep + 1)}{\Gamma(-6 \ep - 1)} \nonumber \\
    & \times
    \left(
        \frac{\GENHYPGF{3}{2}{1, 1 + \ep, 1 + \ep}{1 - \ep, 2 + \ep}{1} }{1+\ep}
        -
        \frac{\GENHYPGF{3}{2}{1, - \ep, 1 + \ep}{1 - \ep, 1 - \ep}{1}}{\ep}
    \right)  .
\end{align}
\end{widetext}

Using the result of the explicit computation of the $1 / \nu$ pole  of ${\bar J}_{2}^{\nu}$
and the result  for ${\bar J}_{1}^{\nu}$ reported in Ref.~\cite{Baranowski:2021gxe},
we find that  they  satisfy the relation shown Eq.~(\ref{eq2.10}).
As we pointed out earlier, this relation
is needed to ensure the smooth $\nu \to 0$ limit of the differential equations.

The above result provides the required boundary condition at $m = \infty$ and allows
us to start solving differential equations numerically. However, it is interesting to
point out that, from   the perspective of the differential equations,
integral ${\bar J}_{2}^{\nu}$ provides an example of a contribution 
proportional to $m^{-2 \nu}$ which, therefore, can mix with the Taylor branch of the required integrals if the $\nu \to 0$
limit is taken first.

Indeed, if we first apply the scaling defined in Eq.~(\ref{eqn:new_region})
to the integration variables in Eq.~(\ref{eqn:j2nutilde_1}),
we find that this integration region leads to an  overall  factor $(1/m)^{2 \nu}$
\begin{align}
\nu {\bar J}_{2}^{\nu}
\sim
 \left(1/m\right)^{2 \nu} \left[ C_{2} + \mathcal{O}(\nu^{1}) \right] \,.
\end{align}

We note that in the limit  $\nu \to 0$, this region looks like a ``normal'' Taylor region $1/m^{0}$
\begin{align}
\nu {\bar J}_{2}^{\nu}
\sim
\left(1/m\right)^{0} C_{2} + \mathcal{O}(\nu^{1}) \,,
\end{align}
and, once the $\nu \to 0$ limit is taken, the two regions cannot be distinguished. This is an illustration
of the problem of mixing between different $m$-branches that we discussed in the previous section; however,
in contrast to the discussion there, our example in this section refers to $m \to \infty$ limit. 
    
The existence of $(1/m)^{2 \nu}$ branch  in the integral $\nu J_2^\nu$ leads to the appearance of similar contributions in
other integrals that $\nu J_2^\nu$ couples to through differential equations. 
For the sake of example,
we  consider the differential equation
for the so-called $J_{a}^{\nu}$ sector, cf. Eq.~(\ref{eq14a}).
The differential equations take the following form 
\begin{widetext}
\begin{align}
        \frac{\partial}{\partial m^{2}}
        \begin{pmatrix}
            J_{a_{1}}^{\nu}
            \\
            J_{a_{2}}^{\nu}
        \end{pmatrix}
        & =
        \begin{pmatrix}
            -\frac{2\ep}{m^2} - \frac{2(1+2\ep)}{4m^2+1} &
            \frac{1}{4 m^{2}+1}
            \\ 
            \frac{2 \ep (1+2 \ep)}{m^{2}} - \frac{8 \ep (1+2 \ep)}{4m^{2}+1} &
            \frac{1+4\ep}{m^{2}} - \frac{28\ep}{4m^{2}+1}
        \end{pmatrix}
        \begin{pmatrix}
            J_{a_{1}}^{\nu}
            \\ 
            J_{a_{2}}^{\nu}
        \end{pmatrix}
        -
        \begin{pmatrix}
            \frac{(1+4\ep)}{(1+6\ep)} \left( \frac{1}{m^{2}} - \frac{4}{4 m^{2}+1} \right)
            \\[0.6em]
            \frac{(1+4\ep)}{(1+6\ep)} \left( \frac{4\ep}{m^{2}} - \frac{16\ep}{4 m^{2}+1} \right)
        \end{pmatrix}
        \nu {\bar J}_{2}^{\nu}
        \label{eqn:deq_Jnu1_a}
        \\
        & \phantom{= {}} +
        \text{contributions from other $J^{\nu}$ integrals.} \nonumber
\end{align}
\end{widetext}
It follows  from the above equation that the integral $\nu {\bar J}_{2}^{\nu}$ plays a role of an inhomogeneous
contribution to the differential equation that  $J_{a_{1}}^{\nu}$ and $J_{a_{2}}^{\nu}$ satisfy.  In fact,
analyzing the homogeneous terms of the above equation in the $m \to \infty$ limit, we find that
we do not need to compute the  boundary conditions for these integrals and that the solution
of the differential equation in this limit are  obtained by integrating the  inhomogeneous part. 
We obtain 
\begin{align}
    J_{a_{1}}^{\nu}
    & =
    m^{-2} \left( \frac{4 \ep + 1}{4 (6 \ep + 1)} C_{2} \right)
    +
    \dots
    ,
    \\
    J_{a_{2}}^{\nu}
    & =
    m^{-2} \left( \frac{\ep (4 \ep + 1)}{6 \ep + 1} C_{2} \right)
    +
    \dots
    ,
\end{align}
where $\nu \to 0$ limit has already been taken where appropriate and dots stand for other contributions,
including homogeneous and
inhomogeneous ones. 

We emphasize one more time that  from the point of view of the differential equation,
the contribution shown in the above equation scales as 
$\sim (1/m)^{2 \nu}$
but, once the $\nu \to 0$ limit is taken it becomes indistinguishable from a regular Taylor part
of the integral. If a similar situation occurs
at $m=0$, we should have identified and removed all the $m^{n_4\nu}$ regions in all integrals since
the correct sequence of limits that is needed is $m \to 0, \nu \to 0, { \ep \to 0}$.  However,
as we mentioned earlier, an analysis of the differential equation at $m \to 0$ leads to the conclusion
that there are no  eigenvalues that vanish if $\nu \to 0$ limit is taken so that  the
problem described above does not   occur.

\section{Checks}
\label{sect:4}

Given the highly unusual nature of the integrals that need to be computed to obtain the
zero-jettiness soft function and the complex interplay of the various infra-red regulators
it is important to perform as many checks as possible to ensure correctness of the result. 

The most comprehensive check  that can be performed is the numerical computation of all $m$-dependent
integrals which  appear in the differential equations, as well as their derivatives. 
We constructed Mellin-Barnes representation of the relevant
integrals using  public programs \texttt{MB}~\cite{Czakon:2005rk} and \texttt{MBresolve}~\cite{Smirnov:2009up}
for this purpose. We also used the program 
\texttt{pySecDec}~\cite{Borowka:2017idc,Borowka:2018goh} as an alternative for the numerical computation.
Using these programs, we have
computed all integrals that
appeared in the differential equations at
finite values of $m$ and checked them against numerical  solutions of the differential equations. 

Next, we compared the solutions of the differential equations at $m=0$ and $\nu = 0$
with the results of the  direct numerical computation.  Unfortunately, although this can be done
for some integrals that contribute to $S_d$, there are many integrals for which
the numerical integration becomes next to impossible.  To enlarge the set of integrals at $m=0$ that
can be checked,   we have derived linear relations
between various integrals at $m=0$ using the integration-by-parts identities and checked that integrals
obtained from $m$-dependent differential equations and extrapolated to $m=0$, satisfy them.

\section{Results}
\label{sect:5}

Solving differential equations and separating the Taylor branch 
 at $m=0$, we obtain numerical result for the integral of the
function $\omega_{n\bar{n}}^{3,(d)}$. Since we can determine  the solution of the differential equation to, essentially, 
arbitrary precision,  we can try to obtain the analytic result for $S_d$ by fitting the numerical results to a linear
combination of various transcendental and rational numbers.
By making use of the  PSLQ~\cite{ferguson1992polynomial} and LLL~\cite{lenstra1982factoring} algorithms, 
and choosing appropriate basis of transcendental numbers~\cite{Henn:2015sem},
we find the following  result
\begin{widetext}
\begin{align}
& S_d = \int {\rm d} \Phi^{nnn}_{\theta \theta \theta} \, \omega^{(3),d}_{n\bar{n}}(k_1,k_2,k_3) =
\frac{12}{\ep^5}
+
\frac{142}{3 \ep^4}
+
\frac{1}{\ep^3} \left( \frac{46 \pi^2}{3} + \frac{628}{3} \right)
+
\frac{1}{\ep^2} \left( 196 \zeta_3 + \frac{650 \pi^2}{9} + \frac{18161}{27} \right)
\nonumber
\\
&
+
\frac{1}{\ep} \left( \frac{397 \pi^4}{45} + 1380 \zeta_3 + \frac{6808 \pi^2}{27} + \frac{165323}{81} \right)
+
\left(
    8982 \zeta_5
    -
    \frac{2146 \zeta_3 \pi^2}{3}
    +
    \frac{191 \pi^4}{9}
    +
    4224 \xLi{4}{\frac{1}{2}}  
\right.
\nonumber
\\
&
\left.
    {}
    +
    3696 \zeta_3 \Ln{2}
    -
    176 \pi^2 \Lnp{2}{2}
    +
    176 \Lnp{4}{2}
    +
    \frac{46184 \zeta_3}{9}
    +
    \frac{66614 \pi^2}{81}
    +
    96 \Ln{2}
    +
    \frac{413971}{81}
\right)
\nonumber
\\
&
+
 \ep
\left(
2304 \MZV{-5, -1} 
-4464 \zeta_{5} \Ln{2}
-8380 \zeta_{3}^2
+\frac{46934 \pi^6}{2835}
-6336 G_{R}(0,0,r_{2},1,-1)
\right.
\nonumber
\\
& \phantom{= {}}
-6336 G_{R}(0,0,1,r_{2},-1)
-3168 G_{R}(0,0,1,r_{2},r_{4})
-6336 G_{R}(0,0,r_{2},-1) \Ln{2}
+\frac{324215 \zeta_{5}}{3}
\label{eqn:res_soft_d}
\\
& \phantom{= {}}
-45056 \xLi{5}{\frac{1}{2}}  
-45056 \xLi{4}{\frac{1}{2}} \Ln{2}  
+176 \CFcl{4}{\frac{\pi}{3}} \pi 
-1056 \zeta_{3} \xLi{2}{\frac1{4}}  
-\frac{9634 \zeta_{3} \pi^2}{3}
\nonumber
\\
& \phantom{= {}}
-21824 \zeta_{3} \Lnp{2}{2}
+2112 \zeta_{3} \Ln{2} \Ln{3}
-1584 \CFclp{2}{2}{\frac{\pi}{3}} \Ln{3} 
-\frac{4400 \CFcl{2}{\frac{\pi}{3}}\pi^3}{27} 
+\frac{88 \pi^4 \Ln{2}}{45}
\nonumber
\\
& \phantom{= {}}
-\frac{616 \pi^4 \Ln{3}}{27}
+\frac{11264 \pi^2 \Lnp{3}{2}}{9}
-\frac{22528 \Lnp{5}{2}}{15}
+8576 \xLi{4}{\frac{1}{2}} 
+7504 \zeta_{3} \Ln{2}
+\frac{4646 \pi^4}{27}
\nonumber
\\
& \phantom{= {}}
-\frac{1072 \pi^2 \Lnp{2}{2}}{3}
+\frac{1072 \Lnp{4}{2}}{3}
+\frac{496592 \zeta_{3}}{27}
-32 \pi^2 \Ln{2}
+\frac{587380 \pi^2}{243}
-384 \Lnp{2}{2}
+832 \Ln{2}
\nonumber
\\
& \phantom{= {}}
+\frac{7857076}{729}
+
\left.
    \sqrt{3}
    \left(
        192 \, \Im \Big\{\!\xLi{3}{\frac{\exp(\ci\pi/3)}{2}}\!\Big\} 
        +160 \CFcl{2}{\frac{\pi}{3}} \Ln{2} 
        -16 \pi \Lnp{2}{2}
        -\frac{560 \pi^3}{81}
   \right)
\right)
\nonumber
\\
& \phantom{= {}}
+
\ORD{\ep^2} \,, \nonumber
\end{align}
\end{widetext}
where $\zeta_{-5, -1} \approx -0.029902$ is a multiple zeta value,
and $\text{Cl}_{n}(x)$ are Clausen functions.
$G_{R}(a_{1}, \ldots, a_{w})$ is the real part of the multiple polylogarithm $G(a_{1}, \ldots, a_{w}; z)$ evaluated at $z = 1$~\cite{Henn:2015sem}
\begin{align}
    G_{R}(a_{1}, \ldots, a_{w})
    =
    \Re \{
    G(a_{1}, \ldots, a_{w}; 1)
    \}
    \,.
\end{align}
Finally, $r_{2} = \exp (-\ci\pi / 3)$ and $r_{4} = \exp (-\ci2 \pi / 3)$.  We note that we have computed
the master integrals to more than two thousand digits to  check the validity of the analytic result.

Having obtained the result for the integral of the function $\omega_{n\bar{n}}^{3,(d)}$ we are now in position to present the
complete result for the same-hemisphere \emph{three-gluon}-emission contribution to the N3LO soft function.
To this end, we write
\begin{widetext}
\begin{align}
        S^{n n n}
        & =
        \int {\rm d} \Phi^{nnn}_{\theta \theta \theta} \left| \boldsymbol{J}(k_1,k_2,k_3) \right|^2
        =
        \tau^{-1 - 6 \ep}
        \frac{N_{\ep}^{3}}{3 !}
        \left[
            C_{a}^{3}
            S^{n n n}_{1 + 1 + 1}
            +
            C_{a}^{2} C_{A}
            S^{n n n}_{1 + 2}
            +
            C_{a} C_{A}^{2}
            S^{n n n}_{3}
        \right]
        ,
\label{eqn:softF_def}
\end{align}
\end{widetext}
where we re-introduced the dependence on $\tau$,
recovered the symmetry factor $1 / 3 !$ and the
normalization factor $N_{\ep}^{3}$ (cf. Eq.~(\ref{eq2a})), and  split the integral into three color
factors following Eq.~(7.10) in Ref.~\cite{Catani:2019nqv}.
We also note that $C_{a} = C_{F,A}$ for the quark (gluon) soft function, 
respectively.

The computation of the Abelian
contributions $S^{n n n}_{1 + 1 + 1}$ and $S^{n n n}_{1 + 2}$ is described in Appendix~\ref{sect:app2_abel}.
We obtain the maximally non-Abelian contribution $S^{n n n}_{3}$ by adding  results obtained
in Ref.~\cite{Baranowski:2021gxe} and the result of this paper given in Eq.~(\ref{eqn:res_soft_d}). We find
\begin{widetext}
\begin{align}
    S^{n n n}_{1 + 1 + 1}
    & =
    \frac{48 \, \Gamma^{3} (1-2 \ep)}{\ep^{5} \Gamma (1-6 \ep)}
    \label{eqn:softF_111}
    ,
    \\
    S^{n n n}_{1 + 2}
    & =
    -\frac{9\, \Gamma (1-4 \ep) \, \Gamma (1-2 \ep)}{\ep^{2} \Gamma (1-6 \ep)}
    \times \bigg[
    \frac{8}{\ep^3} + \frac{44}{3\ep^2} + \frac1{\ep} \left( \frac{268}{9} - 8 \zeta_2 \right) + \left( \frac{1544}{27} + \frac{88}{3} \zeta_2 - 72 \zeta_3 \right)
    \nonumber \\
    & \phantom{= {}}
    + \ep \left( \frac{9568}{81} + \frac{536 \zeta_2}{9} + \frac{352}{3} \zeta_3 - 300 \zeta_4 \right) + \ep^2 \left( \frac{55424}{243} + \frac{3520 \zeta_2}{27} + \frac{2144\zeta_3}{9} + 352 \zeta_4 + 96 \zeta_2 \zeta_3 - 1208 \zeta_5  \right)
    \nonumber \\
    & \phantom{= {}}
    + \ep^3 \left( \frac{297472}{729} + \frac{22592\zeta_2}{81} + \frac{14080\zeta_3}{27} + \frac{2144}{3} \zeta_4 - \frac{4576}{3} \zeta_2 \zeta_3 + 3696 \zeta_5 + 424 \zeta_3^2 - 3596 \zeta_6 \right) + \ORD{\ep^4}
    \bigg]
    \label{eqn:softF_12}
    ,
    \\
    S^{n n n}_{3}
    & =
    \frac{24}{\ep^5}
    +
    \frac{308}{3 \ep^4}
    +
    \frac{1}{\ep^3} \left( -12 \pi^2 + \frac{3380}{9} \right)
    +
    \frac{1}{\ep^2} \left( -1000 \zeta_3 + \frac{440 \pi^2}{9} + \frac{10048}{9} \right)
    \nonumber \\
    & \phantom{= {}}
    +
    \frac{1}{\ep} \left( -\frac{2377 \pi^4}{45} + \frac{440 \zeta_3}{3} + \frac{7192 \pi^2}{27} + \frac{253252}{81} \right)
    \nonumber \\
    & \phantom{= {}}
    +
    \left(
        -
        28064 \zeta_5
        +
        \frac{1972 \zeta_3 \pi^2}{3}
        -
        \frac{638 \pi^4}{15}
        +
        4224 \xLi{4}{\frac{1}{2}}  
        +
        3696 \zeta_3 \Ln{2}
        -
        176 \pi^2 \Lnp{2}{2}
        +
        176 \Lnp{4}{2}
    \right.
    \nonumber \\
    & \phantom{= {}}
        +
    \left.
        \frac{13208 \zeta_3}{3}
        +
        \frac{78848 \pi^2}{81}
        +
        96 \Ln{2}
        +
        \frac{1925074}{243}
    \right)
    \nonumber \\
    & \phantom{= {}}
   +
    \ep
    \left(
    2304 \, \MZV{-5, -1} 
    -4464 \zeta_{5} \Ln{2}
    +25784 \zeta_{3}^2
    -\frac{67351 \pi^6}{567}
    -6336 G_{R}(0,0,r_{2},1,-1)
    \right.
    \nonumber \\
    & \phantom{= {}}
    -6336 G_{R}(0,0,1,r_{2},-1)
    -3168 G_{R}(0,0,1,r_{2},r_{4})
    -6336 G_{R}(0,0,r_{2},-1) \Ln{2}
    +\frac{268895 \zeta_{5}}{3}
    \nonumber \\
    & \phantom{= {}}
    -45056 \xLi{5}{\frac{1}{2}}  
    -45056 \xLi{4}{\frac{1}{2}} \Ln{2}  
     +176 \CFcl{4}{\frac{\pi}{3}} \pi 
     -1056 \zeta_{3} \xLi{2}{\frac1{4}}  
    -3982 \zeta_{3} \pi^2
    \nonumber \\
    & \phantom{= {}}
    -21824 \zeta_{3} \Lnp{2}{2}
    +2112 \zeta_{3} \Ln{2} \Ln{3}
    -1584 \CFclp{2}{2}{\frac{\pi}{3}} \Ln{3} 
    -\frac{4400 \CFcl{2}{\frac{\pi}{3}}\pi^3}{27} 
    +\frac{88 \pi^4 \Ln{2}}{45}
    \nonumber \\
    & \phantom{= {}}
    -\frac{616 \pi^4 \Ln{3}}{27}
    +\frac{11264 \pi^2 \Lnp{3}{2}}{9}
    -\frac{22528 \Lnp{5}{2}}{15}
    +8576 \xLi{4}{\frac{1}{2}} 
    +7504 \zeta_{3} \Ln{2}
    +\frac{4174 \pi^4}{27}
    \nonumber \\
    & \phantom{= {}}
    -\frac{1072 \pi^2 \Lnp{2}{2}}{3}
    +\frac{1072 \Lnp{4}{2}}{3}
    +\frac{554032 \zeta_{3}}{27}
    -32 \pi^2 \Ln{2}
    +\frac{730378 \pi^2}{243}
    -384 \Lnp{2}{2}
    +832 \Ln{2}
    \nonumber \\
    & \phantom{= {}}
    +\frac{1408681}{81}
    +
    \left.
        \sqrt{3}
        \left(
            192 \, \Im \Big\{\!\xLi{3}{\frac{\exp(\ci\pi/3)}{2}}\!\Big\} 
            +160 \CFcl{2}{\frac{\pi}{3}} \Ln{2} 
            -16 \pi \Lnp{2}{2}
            -\frac{560 \pi^3}{81}
       \right)
    \right)
    +
    \ORD{\ep^2}
    .
\end{align}
\end{widetext}

\section{Conclusions}
\label{sect:6}

In this paper, we have discussed the computation of the same-hemisphere
three-gluon-emission contribution to the zero-jettiness soft function at N3LO
in perturbative QCD. We have used the approach   of Ref.~\cite{Baranowski:2021gxe},
which allows us  to apply integration-by-parts technology  and the method of differential equations
to phase-space integrals that contain Heaviside functions. While the appearance of integrals
that are not regulated dimensionally requires an analytic regulator and thus complicates
the use of differential equations, we have described a way to bypass this problem in an efficient way.

Finally, we note that the missing kinematic configuration, in which one of the three gluons
is emitted into the opposite hemisphere can be computed in a similar fashion. Once a complete
result for the three-gluon-emission contribution is known,
the contribution that arises from the emission of a soft $q\bar{q}$-pair and  a soft gluon can be computed in a straightforward way. Similarly, we expect that
virtual corrections to the double real emissions can be dealt with using the same method. 
We leave both problems to future investigations.

{\bf Acknowledgments} This
research was partially supported by the Deutsche
Forschungsgemeinschaft (DFG, German Research Foundation)
under grant 396021762-TRR 257 and by Karlsruhe School
of Particle and Astroparticle physics (KSETA). MD is  supported by the
Excellence Cluster \textsc{ORIGINS} funded by the
Deutsche Forschungsgemeinschaft (DFG, German Research Foundation)
under Germany's Excellence Strategy - EXC-2094 - 390783311 and
by the ERC Starting Grant 949279 \textsc{HighPHun}.

\appendix
\section{Computation of abelian contributions}
\label{sect:app2_abel}
In this appendix we describe how to compute Abelian contributions to the zero-jettiness soft function, i.e.~the first two terms in Eq.~(\ref{eqn:softF_def}). The first, so-called fully-Abelian contribution $S_{1+1+1}^{nnn}$ reads
\begin{align}
\label{eqn:softF_111_def}
S_{1+1+1}^{nnn} =
\int {\rm d} \Phi^{nnn}_{\theta \theta \theta} \, \omega_{n\bar{n}}^{(1)}(k_1) \, \omega_{n\bar{n}}^{(1)}(k_2) \, \omega_{n\bar{n}}^{(1)}(k_3) \,,
\end{align}
where~\cite{Catani:2019nqv}
\begin{align}
\omega_{n\bar{n}}^{(1)}(q) = \frac{4}{(n q) \, (\bar{n}q)}  \,.
\end{align}
Thanks to its fully-factorized structure, the integral in Eq.~(\ref{eqn:softF_111_def})
is straightforward to compute. Using Sudakov variables $\alpha_i$ and $\beta_i$, we find
\begin{align}
  S_{1+1+1}^{nnn} & = 64 \int \limits_{0}^{\infty}
  \left( \PROD{i=1}{3} \DIFFL \alpha_i \DIFFL  \beta_i \, \left( \alpha_i \beta_i \right)^{-1-\ep} \HT{\alpha_i-\beta_i} \right)
   \nonumber \\
   & \times \DD{1-\beta_{123}}
   = \frac{64}{\ep^3} \, \frac{\GFP{3}{-2\ep}}{\GF{-6\ep}} \,.
\end{align}
Making the $1/\ep$-poles in the above equation explicit yields Eq.~(\ref{eqn:softF_111}).

The second Abelian contribution reads
\begin{align}
\label{eqn:softF_12_def}
S_{1+2}^{nnn}  ={} & \int {\rm d} \Phi^{nnn}_{\theta \theta \theta} \Bigg [  \omega_{n\bar{n}}^{(1)}(k_1) \, \omega_{n\bar{n}}^{(2)}(k_2,k_3) \nonumber \\
& + (k_1\leftrightarrow k_2)  \nonumber + (k_1\leftrightarrow k_3) \Bigg ] \nonumber \\
  ={} & 3 \int {\rm d} \Phi^{nnn}_{\theta \theta \theta} \left[  \omega_{n\bar{n}}^{(1)}(k_1) \, \omega_{n\bar{n}}^{(2)}(k_2,k_3) \right] \,.
\end{align}
We write it as
\begin{align}
\label{eqn:softF_12_def_lbl}
& S_{1+2}^{nnn}  = 3 N_\ep^{-1} \int [{\rm d} k_1] \HT{k_1\bar{n} - k_1 n} \, \omega_{n\bar{n}}^{(1)}(k_1) \nonumber \\
& \times N_\ep^{-2} \int \left( \PROD{i=2}{3} [{\rm d} k_i] \HT{k_i\bar{n} - k_i n} \right)  \nonumber \\
& \times \DD{1-k_{123}n} \omega_{n\bar{n}}^{(2)}(k_2,k_3) \,.
\end{align}

The inner integral in Eq.~(\ref{eqn:softF_12_def_lbl}) over $[{\rm d} k_2][{\rm d} k_3]$ can be obtained from the same-hemisphere double-real gluon emission contribution to the NNLO soft function. We find
\begin{align}
\label{eqn:softF_111_res}
& S_{1+2}^{nnn}  = 3 \, \, \INT{0}{\infty} \DIFFL \alpha_1 \DIFFL \beta_1 \HT{\alpha_1-\beta_1} \frac{4 (\alpha_1\beta_1)^{-\ep}}{\alpha_1\beta_1} \nonumber
\\
& \times C^{nn}_2 (1-\beta_1)^{-1-4\ep} = \frac{12}{\ep} \frac{\GF{-4\ep}\GF{-2\ep}}{\GF{-6\ep}} \, C^{nn}_2 \,,
\end{align}
where the factor $C^{nn}_2$ can be extracted from Refs.~\cite{Baranowski:2020xlp,Baranowski:2021gxe}. Upon doing so, we obtain the result displayed in Eq.~(\ref{eqn:softF_12}).

\bibliography{misc/bib.bib}

\end{document}